# Information Hiding and Attacks : Review

Richa Gupta
*Department of Computer Science*
*University of Delhi*
*India*

**ABSTRACT :** *Information Hiding is considered very important part of our lives. There exist many techniques for securing the information. This paper briefs on the techniques for information hiding and the potential threats to those methods. This paper briefs about cryptanalysis and stegananlysis, two methods for breaching into the security methods.*

## 1. INTRODUCTION

Data or information is very crucial to any organization or any individual person. None of us likes our conversation being overheard as it contains the potential of being misused. Same is the case with the data of any organization or of any person. The exchange of data among two potential parties must be in done in a secured method so as to avoid any tampering. Two types of threats exists during any information exchange. The unintended user who may try to overhear this conversation can either tamper with this information to change its original meaning or it can try to listen to the message with intention to decode it and use it to his/her advantage. Both these attacks violated the confidentiality and integrity of the message passed.

Providing intended access and avoiding unintended access is a very challenging task. Information hiding has been since long time. In past, people used hidden pictures or invisible ink to convey secret information [7,8].

## 2. TECHNIQUES FOR INFORMATION HIDING

There are three major data hiding techniques popular: watermarking, cryptography and steganography

**Watermarking** - A watermark is a recognizable image or pattern that is impressed onto paper, which provides evidence of its authenticity [9, 10]. Watermark appears as various shades of lightness/darkness when viewed in transmitted light. Watermarks are often seen as security features to banknotes, passports, postage stamps and other security papers. Digital watermarking is an extension of this concept in the digital world . A watermarking system's primary goal is to ensure robustness, i.e, it should be impossible to remove the watermark without tampering the original data [8].

**Cryptography** - Cryptography is an art of transforming data into an unreadable format called cipher text. The receiver at other side, deciphers or decrypt the message into plain text. Cryptography provides data confidentiality, data integrity, authentication and non-repudiation. Confidentiality is limiting access or placing restriction on certain types of information. Integrity is maintaining and assuring the accuracy of data being delivered, i.e, information contains no modification, deletion etc. Authentication ensures the identity of sender and receiver of the information. Non-repudiation is the ability to ensure that the sender or receiver cannot deny the authenticity of their signature on the sending information that they originated [8].

**Steganography** - Steganography is a practice of hiding/concealing the message, file, image within other message, file or image. The word steganography is of Greek origin and means "covered writing" or "concealed writing" [11]. In other words, it is the art and science of communicating in a way which hides the existence of the communication. The goal is to hide messages inside other harmless messages in a way that does not allow enemy to even detect that there is a second message present [3]. Steganography focuses more on high security and capacity. Even small changes to stego medium can change its meaning. Steganography masks the sensitive data in any cover media like images, audio, video over the internet [8].

### 3. ATTACKS ON INFORMATION HIDING

The attacks on security systems aim to find weakness in information hiding techniques. This is also referred to as breaking. One of the most common example of breaking the security code is brute force approach. To break the security lock of 3-digits, it simply needs 1000 combinations. There are two terms for finding the weakness and trying to break-through the code: cryptanalysis and steganalysis.

**Cryptanalysis**

Cryptanalysis is the study of analyzing information systems in order to study the hidden aspects of the systems. In other words, it is the art of deciphering encrypted communication without knowing the proper keys. It is used to breach cryptographic security systems and to gain access to the encrypted messages. Breaching the security in this way involves knowing how the system works and finding a secret key. Cryptanalysis is the attempt to circumvent the security of various types of cryptographic algorithms and protocols. Types of cryptanalysis attacks:

1. Cipher text only attack – in this, the attacker has access only to a set of ciphertexts. The aim is to deduce plaintexts maybe by making assumptions and guesses.
2. Known plain-text attack - In this, the cryptanalyst has knowledge of a portion of the plaintext from cipher text. The attempt is to deduce key to decrypt the rest of the ciphertext [2,3].
3. Chosen-plaintext attack – this is also known as chosen-cipher text attack or differential cryptanalysis. In this, the cryptanalyst has the ability to choose plaintexts arbitrarily to be encrypted and obtain the corresponding ciphertexts. The cryptanalyst aims to deduce the key by comparing the entire ciphertext with the original plaintext. RSA encryption technique is prone to this type of attack [2,3].
4. Cipher-text only analysis – In this, the cryptanalyst has no knowledge of the plaintext and must work only from ciphertext. It requires guesswork to know what the message can be. Any type of prior knowledge about ciphertext, the sender or the topic in general can be helpful [2].
5. Man-in-the-middle attack – This attack involves tricking individuals into surrendering their keys. When two parties are exchanging their keys for secure communication, an adversary positions himself in between them. He intercepts the signal sent from one side to other, and performs a key exchange separately with both the parties. Thus they both will end up using a different key, known to adversary. The cryptanalyst thus can decrypt the signals between the parties. He can decrypt the communication from party 1 with the key he shared with him, and resends the message by encrypting it with key of other party to other side. Thus both the parties will think they are communicating securely, but the cryptanalyst is hearing everything [3]. This type of attack can be defeated by using hash functions.

6. Timing or differential power analysis – this technique is used to gain information about key computations used in the encryption algorithm and other functions pertaining to security. The technique measures the differences in electrical consumption over a period of time when a microchip performs a function to secure information.

7. Attack against or using the underlying hardware – this type of attack involves the use of mobile crypto devices that aim at the hardware implementation of the cryptosystem. The attacks use the data from very fine measurements of the crypto device doing, say, encryption and compute key information from these measurements. The basic ideas are then closely related to those in other correlation attacks. For instance, the attacker guesses some key bits and attempts to verify the correctness of the guess by studying correlation against her measurements [3].

**Steganalysis**

Steganalysis is the discovery of the existence of hidden information, hidden using steganography. This is analogous to cryptanalysis and cryptography. The goal of steganalysis is to identify suspected packages, to determine whether they have a message encoded into them, and to try and gain access to that message [5]. It differs from cryptanalysis in the sense that the existence of message is obvious in cryptanalysis, that is one knows that the signal contains encrypted data. Whereas in case of steganalysis, the steganalyst starts with a pile of suspect data and then try to determine whether it contains encrypted message and then retrieving the message [4].

1. Stego-only attack – in this type of attack, only the stego media (i.e the medium containing hidden data) is available for analysis [6].
2. Known carrier attack – in this type of attack, the steganalyst has access to both the target object which is used for hiding information and the stego object that contains the hidden information. The stego media or stego object is compared with the cover object and the differences are detected. For example: the original image and the image containing the hidden information are available and compared to deduce the message [5].
3. Known message attack – in this, the original message prior to embedding in the carrier is known. This attack is the analysis of known patterns that correspond to hidden information. This type of analysis can help against attacks in the future. Even with the message available, this type of attack may be very difficult and considered same as stego-only attack.

4. Chosen stego attack – in this attack, the algorithm used for hiding information and the stego object, that is the final hidden file is known and available for analysis [6].
5. Chosen message attack – in this, the steganalyst generates a stego object from some steganography algorithm of a chosen message. The goal is to search for the corresponding patterns in the stego-object that may be helpful for specific steganography tools and algorithms.
6. Known stego attack – in this type of attack, the steganography algorithm, the original mesia file and the stego object is known, that is, all the components are available for analysis [5].

## 4.    CONCLUSION

In this paper, we have provided in brief the methods for hiding information. Also, the potential threats to those security methods are reviewed in detail. The two methods cryptanalysis and steganalysis has been discussed in detail.